\documentclass[conference]{IEEEtran}
\IEEEoverridecommandlockouts

\usepackage{cite}
\usepackage{amsmath,amssymb,amsfonts}
\usepackage{algorithmic}
\usepackage{graphicx}
\usepackage{textcomp}
\usepackage{xcolor}
\usepackage{hyperref} 
\usepackage[misc]{ifsym}
\usepackage{tcolorbox}
\usepackage{multirow}
\usepackage{booktabs}
\usepackage{caption} 
\usepackage{mdframed}

\hypersetup{
  colorlinks=true,
  linkcolor=blue,
  citecolor=blue,
  urlcolor=blue}

\def\BibTeX{{\rm B\kern-.05em{\sc i\kern-.025em b}\kern-.08em
    T\kern-.1667em\lower.7ex\hbox{E}\kern-.125emX}}
\begin{document}

\title{Exploring the Reasoning Depth of Small Language Models in Software Architecture: A Multidimensional Evaluation Framework Towards Software Engineering 2.0\\
}

\author{\IEEEauthorblockN{Ha Vo\textsuperscript{\textsection}, Nhut Tran\textsuperscript{\textsection}, Khang Vo\textsuperscript{\textsection}, Phat T. Tran-Truong$^{\text{(\Letter)}}$ \href{https://orcid.org/0000-0003-3199-6333}{\includegraphics[scale=0.004]{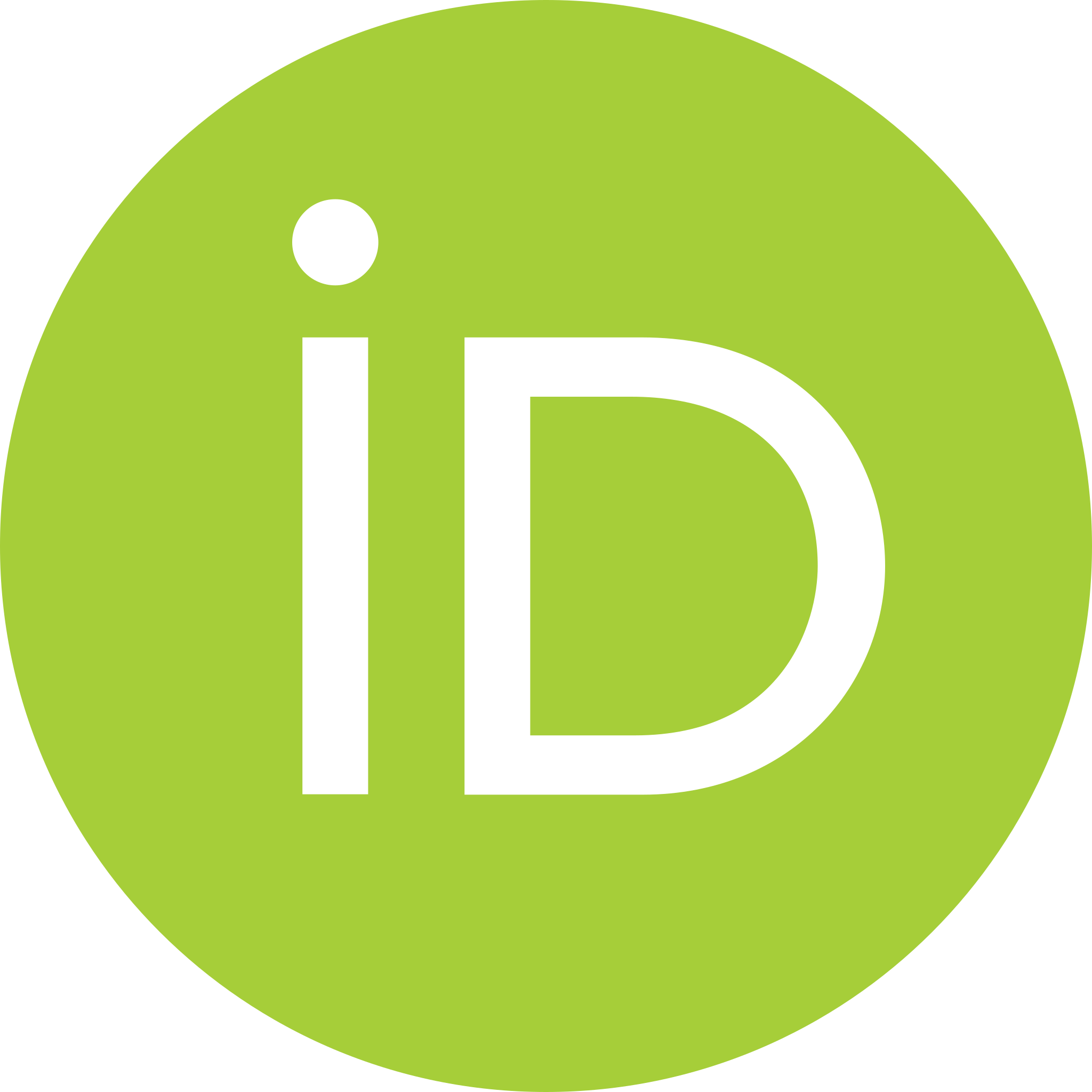}} }
\IEEEauthorblockA{\textit{Faculty of Computer Science and Engineering} \\
\textit{Ho Chi Minh City University of Technology (HCMUT), VNU-HCM}\\
Ho Chi Minh City, Vietnam \\
\{ha.voxuancs22, nhut.trannov25th, khang.vohuu,  phatttt\}@hcmut.edu.vn}
\and
\IEEEauthorblockN{Son Ha}
\IEEEauthorblockA{\textit{The Business School} \\
\textit{RMIT University}\\
Ho Chi Minh City, Vietnam \\
ha.son@rmit.edu.vn}

}

\maketitle
\begingroup\renewcommand\thefootnote{\textsection}
\footnotetext{These authors contributed equally.\\ 
$^{\text{(\Letter)}}$ Corresponding Author: Phat T. Tran-Truong (phatttt@hcmut.edu.vn)
}
\endgroup


\begin{abstract}
 In the era of "Software Engineering 2.0" (SE 2.0), where intelligent agents collaborate with human engineers, Generative AI is advancing beyond code generation into Software Architecture (SA). While Large Language Models (LLMs) demonstrate superior capabilities, computational costs and data privacy concerns drive interest in Small Language Models (SLMs) with fewer than 7 billion parameters. However, the reasoning limits of these resource-constrained models remain unexplored. This study benchmarks 10 state-of-the-art SLMs on Architectural Decision Records generation, introducing a multi-dimensional framework evaluating Technical Compliance and Semantic Diversity. Our empirical results reveal a significant reasoning gap: models above the 3B-parameter threshold demonstrate robust zero-shot capabilities, while sub-2B models show the strongest BERTScore gains from Fine-Tuning, though compliance improvements are not guaranteed. Contrary to assumptions regarding context saturation, Few-Shot prompting serves as a highly effective calibration mechanism for select mid-sized models with short context windows. Furthermore, high semantic diversity in off-the-shelf small models often correlates with hallucination rather than productive exploration. These findings establish a rigorous baseline for deploying sustainable, locally hosted architectural assistants.
\end{abstract}


\section{Introduction}
The field of Software Engineering (SE) is witnessing a paradigm shift toward ``Software Engineering 2.0''\footnote{We adopt the agent-centric definition of SE~2.0 from \cite{h1}, which characterizes it as the collaboration between human engineers and autonomous intelligent agents. This differs from earlier industry usages of the term that emphasize probabilistic programming or neural networks replacing explicit code~\cite{h2}.}, where software development transforms into collaboration between engineers and autonomous intelligent agents\cite{h1}. Recent advances in generative AI have enabled software developers to shift focus toward higher-level design activities: requirements analysis, system design, and Software Architecture (SA)\cite{h2,HouZLYWLLLGW24}, while developers can dedicate low-level tasks such as implementation, code comprehension, debugging, etc. to software agents \cite{yang2024swe,wei2023copiloting,lee2025unidebugger}. SA serves as the backbone of every digital system, requiring abstract thinking, trade-off analysis between quality attributes, and long-term maintainability forecasting. However, this shift prompts a critical question: can generative AI be trusted to support - or even make - architectural decisions that require balancing competing quality attributes and reasoning about complex trade-offs?

Large Language Models (LLMs) have demonstrated exceptional capabilities in generating Architecture Decision Records (ADRs)\cite{h3}. However, LLMs come at significant cost. Their extensive computational resources lead to prohibitively high operational expenses and substantial carbon emissions, making the search for efficient solutions both a moral and economic imperative\cite{h4}. Additionally, architectural documents often contain business secrets and system vulnerabilities, causing organizations to prohibit sending such data to public cloud APIs.

Small Language Models (SLMs)\cite{h5}, typically defined as having fewer than 7 billion parameters, have emerged as a promising solution. They enable on-premise deployment, ensuring data security while reducing latency and inference costs through techniques such as model pruning, knowledge distillation, and low-rank factorization\cite{h6}. While enterprise license agreements can partially mitigate cloud data-sharing concerns, they do not address the recurring per-token operational costs that accumulate during iterative architectural refinement workflows, the inference latency inherent to remote API calls, regulatory constraints in sectors such as defense and healthcare that mandate air-gapped environments, or the environmental footprint of continuous cloud-based inference. Parameter-Efficient Fine-Tuning (PEFT) methods like LoRA make fine-tuning accessible even on consumer-grade hardware, positioning SLMs as a sustainable complement to cloud-based LLMs rather than a wholesale replacement.

Despite SLMs' growing prominence, a notable gap exists in their systematic evaluation for SA tasks. Existing benchmarks like SLM-Bench\cite{h7} evaluate diverse NLP tasks, while code generation benchmarks like HumanEval\cite{chen2021humaneval}, MBPP\cite{austin2021program}, and SWE-bench\cite{h10} primarily assess functional correctness rather than architectural reasoning. Critically, existing ADR generation studies\cite{h3} rely on textual similarity metrics (ROUGE, BLEU), which cannot detect architecturally incorrect but grammatically correct designs.

To address this gap, we systematically evaluate open-weights SLMs in generating ADRs, making three significant contributions:

\textbf{(1) SLM-ArchBench Framework}: A comprehensive evaluation framework moving beyond textual similarity to assess architectural reasoning, probing the parameter threshold where models transition from lexical mimicry to genuine architectural compliance.

\textbf{(2) Domain Adaptation Strategy Comparison}: Systematic comparison of In-Context Learning versus Parameter-Efficient Fine-Tuning to identify optimal deployment configurations for resource-constrained environments.

\textbf{(3) Quality-Diversity Trade-off Analysis}: Empirical investigation of how the inherent tension between output diversity and generation quality-common across generative models-manifests under the parameter constraints of SLMs, establishing practical boundaries between productive architectural exploration and stochastic hallucination.

We evaluate 10 open-weights SLMs across varying parameter sizes, providing a principled basis for deploying sustainable and reliable architectural assistants balancing resource efficiency with real-world applicability in SA.

\section{Related Work}

\subsection{Language Models for SA}

The application of language models to SA represents a natural progression from earlier successes in code generation. Dhar et al. \cite{h3} demonstrated that GPT-4 exhibits noteworthy capabilities in generating Architecture Decision Records (ADRs), while smaller models like Flan-T5-base achieve comparable results with few-shot prompts and fine-tuning. Contemporary research has explored design rationale mining \cite{zhao2024drminer, dhaouadi2024rationale}, knowledge graph construction \cite{palombo2026generate, chen2023autokg}, and domain-specific evaluation resources such as the QuArch dataset \cite{prakash2025quarch}. Despite these advances, the systematic evaluation of SLMs for architectural reasoning remains underexplored, leaving practitioners without clear guidance on whether resource-efficient alternatives can adequately support architectural decision-making in privacy-sensitive or resource-constrained environments.

\subsection{Benchmarks for Code and Reasoning Tasks}

Established benchmarks such as HumanEval \cite{chen2021humaneval}, MBPP \cite{austin2021program}, and SWE-bench \cite{h10} have become standard tools for evaluating code generation capabilities, yet predominantly assess functional correctness at the implementation level, neglecting higher-order architectural reasoning. Recent efforts like SLM-Bench \cite{saip} integrate environmental metrics alongside accuracy across diverse NLP tasks, while multi-agent frameworks \cite{li2025smoa, li2024coevol} have been proposed for iterative evaluation enhancement. However, these approaches remain largely general-purpose, lacking the domain-specific criteria necessary for assessing architectural competence, including anticipating trade-offs, ensuring documentation-implementation alignment, and analyzing change impact propagation \cite{saip}.

\subsection{Evaluation Metrics: Limitations and Advances}

Traditional text similarity metrics such as ROUGE and BLEU exhibit fundamental limitations when applied to technical domains \cite{goodrich2019assessing, grusky2023rogue} - a design decision may achieve high textual similarity while violating core architectural principles. Recognition of these limitations has driven the development of model-based metrics including BERTScore \cite{pan2024humancentered} and COMET \cite{leiter2024towards}, the LLM-as-a-Judge paradigm \cite{gebreegziabher2025metricmate, hu2025training}, and multi-dimensional frameworks like Meta-rater \cite{zhuang2025metarater} and AXCEL \cite{sreekar2024axcel} that integrate multiple quality dimensions. These advances inform our proposed framework, which extends beyond textual similarity to assess semantic diversity, structural compliance, and impact prediction capabilities.

\section{Study design and execution}

\begin{figure*} [t]
    \centering
    \includegraphics[width=0.7\linewidth]{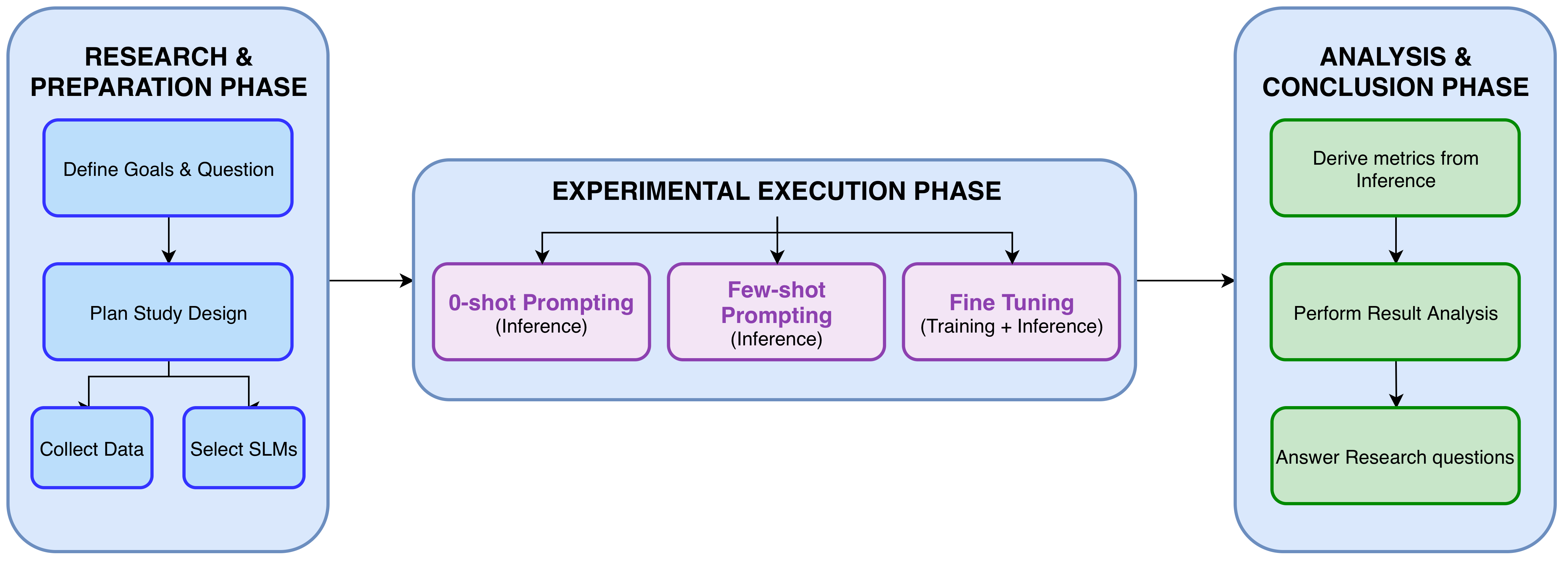}
    \caption{Study Design}
    \label{fig:studydesign}
\end{figure*}

This section documents the design and execution of our empirical study. We begin by defining the research goal (Section \ref{Goal}) and formulating the specific research questions (Section \ref{researchquestions}). Next, we detail the model selection criteria (Section \ref{modelsection}) and the data collection methodology (Section \ref{datacollection}). The study then proceeds to the experimental procedure (Section \ref{procedure}), describing the execution pipeline and generation strategies, finally concluding with the definition of the multi-dimensional evaluation metrics (Section \ref{metrics}). A visualization of the complete experimental workflow is provided in Fig. \ref{fig:studydesign}.
\subsection{Goal} \label{Goal}
The primary goal of this research is to develop and apply a novel evaluation methodology that probes the reasoning depth of SLMs in the domain of SA. While existing benchmarks focus predominantly on code generation or general NLP tasks with surface-level metrics, these approaches assess functional correctness rather than the kind of higher-order reasoning that architectural decision-making demands. ADR generation is therefore selected as the evaluation task: it requires a model to simultaneously analyze contextual constraints, propose a technically sound solution, and justify trade-offs against competing quality attributes. ADRs' well-defined structure and established role as an SA benchmark further ensure systematic, reproducible evaluation with direct comparability to prior work. We acknowledge that real-world ADRs vary widely in analytical depth and architectural insight; we specifically address this variability through strict dataset curation (Section~\ref{datacollection}) and an evaluation protocol that scores technical validity of decision rationale rather than superficial formatting adherence (Section~\ref{metrics}).

This study aims to establish SLM-ArchBench - a systematic framework that quantifies architectural reasoning through semantically and structurally grounded evaluation dimensions - specifically to:

\begin{itemize}
    \item \textbf{Quantify the architectural reasoning capabilities} of state-of-the-art SLMs when faced with complex SA problems.
    \item \textbf{Analyze the trade-offs} between model size and reasoning performance, providing actionable insights for practitioners seeking to deploy AI-assisted architectural tools in resource-constrained or privacy-sensitive environments.
\end{itemize}

We define ``reasoning depth" through three fundamental aspects: the ability to explore a diverse solution space rather than producing superficial variations, strict compliance with established structural rules and design patterns, and accurate prediction of the ripple effects caused by architectural changes. We note that diversity and quality trade-offs are general properties of generative models; our specific contribution is characterizing how these phenomena manifest under the parameter constraints of SLMs, where limited capacity amplifies their practical consequences for architectural reasoning. This multidimensional definition ensures that evaluation reflects not merely linguistic fluency but genuine architectural understanding.
\subsection{Research Questions} \label{researchquestions}
To achieve our goal of benchmarking efficient models for architectural decision support, we address the following four research questions (RQ):

\textbf{RQ1} \textit{How capable are off-the-shelf SLMs in generating semantically accurate and architecturally compliant design decisions?}

This question establishes the baseline capability of resource-constrained models to generate coherent ADRs without prior training. We specifically investigate whether these efficient models possess sufficient pre-trained knowledge to propose technically sound architectural patterns that comply with industry best practices, or if they merely replicate the terminology without grasping the underlying design rationale.

\textbf{RQ2} \textit{Does In-Context Learning improve the architectural compliance and semantic quality of SLMs?}

This question examines the impact of providing limited examples ($k=2$) within the prompt context. We investigate whether the addition of valid ADR examples serves as an effective calibration mechanism to guide the models toward standard architectural patterns, thereby increasing their compliance scores, or if the increased context length introduces noise that degrades reasoning in constrained attention windows.

\textbf{RQ3} \textit{To what extent does domain-specific Fine-Tuning maximize architectural compliance in resource-constrained models?}

This question explores the efficacy of Parameter-Efficient Fine-Tuning (LoRA) as a domain adaptation strategy. We analyze whether fine-tuning is strictly necessary to correct the reasoning deficits and compliance gaps of the smallest models (1B parameters) and whether it offers significant advantages over prompting strategies for larger, more capable architectures.

\textbf{RQ4} \textit{To what extent do SLMs exhibit semantic diversity when proposing solutions to open-ended architectural requirements?}

This question analyzes the breadth of the solution space explored by the models. We seek to differentiate between "productive exploration"-where the model proposes varied but valid trade-offs-and "stochastic variance," where high diversity signals hallucination or lack of compliance. We further investigate how different adaptation strategies (In-Context Learning vs. Fine-Tuning) influence this balance between convergence on compliant solutions and creative exploration.

\subsection{Model Selection} \label{modelsection}
\subsubsection{Selection Criteria and Rationale}
 We define selection criteria ensuring comprehensive coverage while maintaining experimental feasibility. Our model selection adheres to four fundamental principles: (1) \textbf{Parameter Constraint:} Models must contain fewer than 7 billion parameters, aligning with the conventional SLM definition \cite{h5} and representing a practical boundary for consumer-grade hardware deployment; (2) \textbf{Open-Weights Availability:} All models must be publicly accessible with permissive licenses to ensure reproducibility; (3) \textbf{Instruction-Tuned Variants:} We exclusively utilize 
instruction-tuned versions, as ADR generation requires structured 
instruction-following \cite{a4}, reflects practical SE 2.0 deployment 
scenarios \cite{h3}, and benefits from the enhanced format compliance 
and multi-step reasoning capabilities that instruction tuning provides 
\cite{a3, a6}; (4) \textbf{Model Lineage and Training Diversity:} We selected models from diverse providers (Meta, Microsoft, Google, Alibaba, Mistral) to enable comparative analysis of how different training methodologies influence architectural reasoning capabilities.

\subsubsection{Selected Models}
Based on these criteria, we selected 10 instruction-tuned SLMs representing diverse approaches to model design, training data composition, and architectural innovation.

\renewcommand{\arraystretch}{1}
\begin{table}[h]
\centering
\resizebox{0.5\textwidth}{!}{%
\begin{tabular}{|l|l|c|c|l|}
\hline
\textbf{Model Name} & \textbf{Provider} & \textbf{Parameters (B)} & \textbf{Context Length} & \textbf{Release Date} \\
\hline
Llama-3.2-1B & Meta & 1.2 & 131,072 & Sept 2024 \\
\hline
Llama-3.2-3B & Meta & 3.2 & 131,072 & Sept 2024 \\
\hline
Phi-3-mini & Microsoft & 3.8 & 4,096 & Apr 2024 \\
\hline
OLMo-2-1B & Allen Institute (Ai2) & 1.0 & 4,096 & Apr 2025 \\
\hline
OLMo-2-7B & Allen Institute (Ai2) & 7.0 & 4,096 & Nov 2024 \\
\hline
Qwen2.5-1.5B & Alibaba & 1.54 & 32,768 & Sept 2024 \\
\hline
Qwen2.5-3B & Alibaba & 3.09 & 32,768 & Sept 2024 \\
\hline
Gemma-3-1B & Google DeepMind & 1.0 & 32,768 & Mar 2025 \\
\hline
SmolLM2-1.7B & Hugging Face & 1.7 & 8,192 & Nov 2024 \\
\hline
Mistral-7B-v0.3 & Mistral AI & 7.0 & 32,768 & May 2024 \\
\hline
\end{tabular}
}
\end{table}

\subsection{Data Collection} \label{datacollection}
\subsubsection{Dataset Selection Rationale}
We adopt an established dataset specifically designed for evaluating LLMs on ADR generation \cite{h3}, motivated by three considerations. First, it ensures direct comparability with prior work. Second, unlike general NLP datasets, it comprises 95 domain-specific ADRs systematically collected from five GitHub repositories, created by practicing architects following standardized formats. Third, current SLM benchmarks \cite{h7} lack evaluation scenarios reflecting real-world architectural decision-making.

\textbf{Data Characteristics:} The dataset comprises 95 Context-Decision pairs from archane-framework (17), winery (17), joelparkerhenderson (32), cardano (14), and island (15) \cite{h3}. Each ADR follows standard format: Context describes the problem and constraints; Decision articulates the solution and rationale.

\subsubsection{Dataset Limitations and Scope}
We acknowledge that real-world ADRs vary widely in their analytical depth and architectural insight, posing a significant challenge for benchmark construction. To mitigate this variability, we specifically restricted our evaluation to a curated dataset of 95 expert-authored ADRs \cite{h3} rather than conducting large-scale scraping. These ADRs were systematically collected from five GitHub repositories and created by practicing architects following a standardized ``Context'' and ``Decision'' format, ensuring a baseline level of analytical rigor. Furthermore, to ensure we evaluate actual architectural insight rather than superficial structure, our automated compliance evaluator (Section~\ref{metrics}) is strictly instructed to score the technical validity of the decision rationale and its alignment with architectural best practices, rather than mere formatting adherence.

The 95 samples are modest but reflect domain constraints---high-quality ADRs require expert knowledge, making large-scale collection infeasible \cite{h3}. ADRs from open-source projects \cite{h3} may not fully represent enterprise contexts. We converted Markdown to CSV for: (1) seamless integration with data loading libraries, and (2) removing formatting noise for smaller models. Following baseline methodology \cite{h3}, we used 80\% training and 20\% validation splits, with validation data for zero/few-shot experiments and few-shot examples from training data to prevent leakage. All ADRs are from publicly accessible repositories with open-source licenses \cite{h3}.

\subsection{Experimental Procedure} \label{procedure}

To rigorously evaluate the architectural reasoning capabilities of SLMs, we executed a multi-stage experimental pipeline. All models were loaded using 4-bit Normal Float (NF4) quantization. The experiment was conducted in three distinct phases:

\subsubsection{Zero-Shot}

In this baseline setting, we evaluated the models' innate capability to generate ADRs without prior examples. The model was provided solely with the decision context and a specific instruction to generate the decision. The exact prompt template is illustrated in Figure \ref{zero_short_prompt}.

\begin{figure}[h]
\centering
\begin{tcolorbox}[width=0.9\linewidth, boxrule=0.5pt]
You are a SA assistant. Given the following architecture decision context, write a clear and complete Architecture Decision Record (ADR) decision.

Context:
\{target\_context\}

Decision:
\end{tcolorbox}
\caption{Zero-shot prompt}
\label{zero_short_prompt}
\end{figure}

\subsubsection{Few-Shot}

To assess the impact of the few-shot approach (in-context learning), we employed a few-shot prompting strategy with $k=2$ . We selected 2 examples from the training set according to the golden samples in \cite{h3} and added them to the target prompt. The structure of the few-shot prompt is presented in Figure \ref{few_short_prompt}.
\begin{figure}[h]
\centering
\begin{tcolorbox}[width=0.9\linewidth, boxrule=0.5pt]
You are a SA assistant. Given the following architecture decision context, write a clear and complete Architecture Decision Record (ADR) decision.

Context:
\{example\_context\_1\}

Decision:
\{example\_decision\_1\}

Context:
\{example\_context\_2\}

Decision:
\{example\_decision\_2\}

Context:
\{target\_context\}

Decision:
\end{tcolorbox}
\caption{Few-shot prompt}
\label{few_short_prompt}
\end{figure}

\subsubsection{Parameter-Efficient Fine-Tuning (PEFT)}

We specialized the SLMs for the ADR generation task using Low-Rank Adaptation (LoRA), training on the 80\% training split (76 Context-Decision pairs) of the curated ADR dataset described in Section~\ref{datacollection}. Each training sample pairs an architectural context as input with the corresponding expert-authored decision as the target output. We targeted all linear modules within the models using a rank $r=16$, alpha $\alpha = 32$ , and a dropout rate of $0.5$. The models were fine-tuned for 10 epochs using the AdamW optimizer with a learning rate of $2 \times 10^{-4}$ and a linear warmup over the first 10\% of steps. To accommodate memory constraints, we utilized a batch size of 2 with 4 gradient accumulation steps. The complete training data, fine-tuning scripts, and trained adapter weights are available in our replication package.\footnote{\url{https://github.com/LeoTTTPhat/SLM-ArchBench}}

\subsection{Metrics} \label{metrics}
To provide a holistic evaluation of the generated architectural decisions, we utilized a multi-dimensional metric suite covering lexical overlap, semantic accuracy, structural validity, and solution diversity.

\subsubsection{Standard Evaluation Metrics} 

We employed a comprehensive set of NLP metrics to measure the textual similarity between the generated decisions and the ground truth:

BERTScore (F1): Utilized as our primary measure of semantic accuracy, this metric evaluates the contextual embedding similarity between candidate and reference texts, capturing meaning beyond exact word matches.

ROUGE Family (1, 2, L): These recall-oriented metrics quantify the extent to which the generated text covers the content of the ground truth, measuring overlap at the unigram, bigram, and longest-common-subsequence levels.

BLEU: This metric assesses the precision of the generation, penalizing the inclusion of irrelevant or hallucinated tokens that do not appear in the reference text.

METEOR: To account for linguistic variability, this metric evaluates alignment by considering synonyms, stemming, and paraphrasing, providing a more flexible measure of correctness than strict exact-matching.

\subsubsection{Architectural Compliance Score (LLM-as-a-Judge)}

Standard metrics often fail to capture technical correctness or adherence to specific architectural patterns. To address this, we implemented an automated Architectural Compliance Score using \textit{Gemini-2.5-Flash} as an expert evaluator.
The evaluator compares the model's output against the ground truth and assigns a scalar score ($0-100$) based on the technical validity of the decision rationale and its alignment with architectural best practices. The exact prompt and scoring rubric are defined in Figure~\ref{fig:compliance_prompt}.

\begin{figure}[h]
\centering
\begin{tcolorbox}[width=0.9\linewidth, boxrule=0.5pt]
You are an expert SA evaluator.

Given a model's generated architectural decision and the ground truth, rate how well the model's answer aligns with standard architectural patterns and the ground truth decision.
\\
Consider:

- Correctness of architectural approach

- Alignment with the ground truth rationale

- Compliance with architectural best practices (e.g., MVC, microservices, layered architecture)\\

Model Answer:
\{model\_answer\}

Ground Truth:
\{ground\_truth\}\\

Rate the compliance from 0 to 100:

- 0-20: Completely wrong or irrelevant \\
- 21-40: Partially correct but major issues\\
- 41-60: Somewhat aligned but significant gaps \\
- 61-80: Good alignment with minor differences \\
- 81-100: Excellent alignment, equivalent or better\\

Respond with ONLY a number between 0 and 100. No explanation.
\end{tcolorbox}
\caption{Automated Compliance Evaluator Prompt}
\label{fig:compliance_prompt}
\end{figure}

\subsubsection{Semantic Diversity Score}

To quantify the breadth of the solution space explored by the models (RQ4), we moved beyond single-point estimates. For each test input, we generated 3 distinct candidate solutions using nucleus sampling.

We encoded these solutions into dense vector representations to calculate the Semantic Diversity Score, defined as the mean pairwise cosine distance between all candidate solutions for a given input. A higher score indicates that the model proposed architecturally distinct solutions, while a score near zero indicates mode collapse or repetitive generation.

\section{Results}
In this section, we present the findings of our experimentation in line with the research questions that frame this study  (Section \ref{researchquestions}). Tables \ref{tab:zeroshot_result}, \ref{tab:fewshot_result}, \ref{tab:finetune_result} summarize the metric scores obtained from the zero-shot, few-shot, and fine-tuning configurations, respectively, with the best score for each metric highlighted in bold. Our main evaluation measure, BERTScore (F1), is visualized in Figure~\ref{fig:bertscore_f1}.

\subsection{Results RQ1: How capable are off-the-shelf SLMs in generating semantically accurate and architecturally valid design decisions?}

\begin{table*}[t]
\centering

\begin{tabular}{l|ccccc|ccc|cc}
\toprule
\multirow{2}{*}{\textbf{model}} &
  \multirow{2}{*}{\textbf{rouge1}} &
  \multirow{2}{*}{\textbf{rouge2}} &
  \multirow{2}{*}{\textbf{rougeL}} &
  \multirow{2}{*}{\textbf{bleu}} &
  \multirow{2}{*}{\textbf{meteor}} &
  \multicolumn{3}{c|}{\textbf{BERTScore}} &
  \multirow{2}{*}{\textbf{diversity}} &
  \multirow{2}{*}{\textbf{compliance}} \\ 
 & & & & & & \textbf{precision} & \textbf{recall} & \textbf{f1} & & \\ 
\midrule
gemma-3-1b      & 0.171 & 0.026 & 0.107 & 0.006 & 0.161 & 0.78  & 0.833 & 0.805 & 0.397 & 45.421 \\ 
llama-3.2-1b    & 0.181 & 0.032 & 0.108 & 0.015 & 0.173 & 0.802 & 0.839 & 0.82  & 0.417 & 53.947 \\ 
llama-3.2-3b    & 0.192 & 0.032 & 0.112 & 0.016 & 0.186 & 0.809 & \textbf{0.845} & 0.826 & 0.342 & 65.421 \\ 
mistral-7b-v0.3 & \textbf{0.202} & \textbf{0.042} & \textbf{0.117} & \textbf{0.018} & \textbf{0.189} & \textbf{0.813} & 0.844 & \textbf{0.827} & 0.28  & 66.947 \\ 
olmo-2-1b       & 0.18  & 0.029 & 0.102 & 0.017 & 0.177 & 0.809 & 0.843 & 0.825 & 0.298 & 60.474 \\ 
olmo-2-7b       & 0.187 & 0.03  & 0.104 & 0.009 & 0.184 & 0.805 & 0.835 & 0.819 & 0.274 & 62.368 \\ 
phi-3-mini      & 0.173 & 0.029 & 0.104 & 0.013 & 0.171 & 0.801 & 0.837 & 0.818 & 0.499 & 66.421 \\ 
qwen2.5-1.5b    & 0.166 & 0.02  & 0.089 & 0.012 & 0.173 & 0.809 & 0.841 & 0.824 & 0.365 & 56.316 \\ 
qwen2.5-3b      & 0.184 & 0.03  & 0.104 & 0.013 & 0.184 & 0.806 & 0.842 & 0.823 & 0.283 & \textbf{71.737} \\ 
smollm2-1.7b    & 0.165 & 0.033 & 0.098 & 0.021 & 0.168 & 0.799 & 0.834 & 0.815 & \textbf{0.541} & 51.053 \\ 
\bottomrule
\end{tabular}%

\caption{Zero-shot Result}
\label{tab:zeroshot_result}
\end{table*}

To address RQ1, we evaluated the zero-shot baseline of 10 SLMs. Mistral-7b-v0.3 achieved the strongest semantic similarity (BERTScore F1: 0.827), while Qwen2.5-3b led in architectural compliance (71.7\%). Most models above the 3B-parameter threshold achieved compliance scores above 65\%, with OLMo-2-7b (62.4\%) as the notable exception.

In contrast, Gemma-3-1b achieved a competitive BERTScore of 0.805 but recorded the lowest Compliance Score in the cohort at 45.4\%. Similarly, Smollm2-1.7b scored 51.1\% in compliance despite a BERTScore of 0.815, revealing a consistent divergence between lexical and architectural performance in sub-2B models.

\subsection{Results RQ2: Does In-Context Learning (Few-Shot Prompting) improve the architectural correctness and semantic quality of SLMs?}

\begin{table*}[t]
\centering
\begin{tabular}{lccccc|ccc|cc}
\toprule
\multirow{2}{*}{\textbf{model}} &
  \multirow{2}{*}{\textbf{rouge1}} &
  \multirow{2}{*}{\textbf{rouge2}} &
  \multirow{2}{*}{\textbf{rougeL}} &
  \multirow{2}{*}{\textbf{bleu}} &
  \multirow{2}{*}{\textbf{meteor}} &
  \multicolumn{3}{c|}{\textbf{BERTScore}} &
  \multirow{2}{*}{\textbf{diversity}} &
  \multirow{2}{*}{\textbf{compliance}} \\ 
 & & & & & & \textbf{precision} & \textbf{recall} & \textbf{f1} & & \\ 
\midrule
gemma-3-1b      & 0.202 & 0.06  & 0.139 & 0.045 & 0.198 & 0.791 & 0.84  & 0.814 & 0.463 & 46.474 \\ 
llama-3.2-1b    & 0.186 & 0.055 & 0.119 & 0.042 & 0.193 & 0.8   & 0.839 & 0.818 & 0.426 & 44.105 \\ 
llama-3.2-3b    & 0.195 & 0.042 & 0.11  & 0.023 & 0.188 & 0.811 & 0.85  & 0.83  & 0.392 & 57.105 \\ 
mistral-7b-v0.3 & 0.224 & 0.076 & 0.147 & \textbf{0.055} & 0.216 & \textbf{0.823} & 0.85  & \textbf{0.835} & 0.303 & 62.0   \\ 
olmo-2-1b       & 0.2   & 0.068 & 0.131 & 0.042 & 0.207 & 0.81  & 0.844 & 0.826 & 0.36  & 56.211 \\ 
olmo-2-7b       & \textbf{0.229} & 0.071 & 0.14  & 0.049 & 0.211 & 0.819 & \textbf{0.853} & \textbf{0.835} & 0.252 & \textbf{73.632} \\ 
phi-3-mini      & 0.219 & 0.068 & 0.14  & 0.051 & 0.198 & 0.817 & 0.848 & 0.831 & 0.554 & 72.105 \\ 
qwen2.5-1.5b    & 0.172 & 0.022 & 0.092 & 0.016 & 0.169 & 0.807 & 0.843 & 0.824 & 0.439 & 52.053 \\ 
qwen2.5-3b      & 0.203 & 0.063 & 0.134 & 0.043 & \textbf{0.221} & 0.8   & 0.848 & 0.822 & 0.296 & 67.316 \\ 
smollm2-1.7b    & 0.216 & \textbf{0.081} & \textbf{0.148} & 0.05  & 0.214 & 0.815 & 0.843 & 0.828 & \textbf{0.642} & 65.684 \\ 
\bottomrule
\end{tabular}%
\caption{Few-shot Result}
\label{tab:fewshot_result}
\end{table*}

RQ2 investigates the efficacy of In-Context Learning (ICL) by providing two example ADRs in the prompt. ICL produced mixed compliance results: four models improved while six experienced declines (Table~\ref{tab:fewshot_result}).

Among the beneficiaries, the gains were substantial. Phi-3-mini's compliance rose by +5.7 percentage points (pp), and OLMo-2-7b recorded the largest gain at +11.2~pp, achieving the highest few-shot compliance in the cohort.

In contrast, ICL degraded performance in the majority of models. Llama-3.2-1b experienced the sharpest decline ($-$9.8~pp) and Mistral-7b-v0.3 dropped by $-$4.9~pp, suggesting that for models with strong zero-shot priors, the addition of examples introduces noise rather than a useful calibration signal.

\subsection{Results RQ3: To what extent does domain-specific Fine-Tuning maximize architectural correctness in resource-constrained models?}

\begin{table*}[]
\centering
\begin{tabular}{lccccc|ccc|cc}
\toprule
\multirow{2}{*}{\textbf{model}} &
  \multirow{2}{*}{\textbf{rouge1}} &
  \multirow{2}{*}{\textbf{rouge2}} &
  \multirow{2}{*}{\textbf{rougeL}} &
  \multirow{2}{*}{\textbf{bleu}} &
  \multirow{2}{*}{\textbf{meteor}} &
  \multicolumn{3}{c|}{\textbf{BERTScore}} &
  \multirow{2}{*}{\textbf{diversity}} &
  \multirow{2}{*}{\textbf{compliance}} \\ 
 & & & & & & \textbf{precision} & \textbf{recall} & \textbf{f1} & & \\ 
\midrule
gemma-3-1b      & \textbf{0.202} & \textbf{0.042} & \textbf{0.116} & 0.022 & \textbf{0.203} & 0.812 & 0.846 & 0.828 & 0.382 & 47.895 \\ 
llama-3.2-1b    & 0.174 & 0.028 & 0.101 & 0.012 & 0.164 & 0.804 & 0.838 & 0.82  & 0.437 & 53.842 \\ 
llama-3.2-3b    & 0.197 & 0.04  & 0.113 & 0.02  & 0.193 & 0.808 & 0.845 & 0.825 & 0.404 & \textbf{58.263} \\ 
mistral-7b-v0.3 & 0.2   & 0.035 & 0.109 & \textbf{0.024} & 0.198 & 0.813 & 0.85  & \textbf{0.83}  & 0.377 & 54.211 \\ 
olmo-2-1b       & 0.181 & 0.039 & 0.107 & 0.023 & 0.172 & 0.767 & 0.797 & 0.781 & 0.452 & 46.632 \\ 
olmo-2-7b       & 0.195 & 0.034 & 0.11  & 0.016 & 0.178 & \textbf{0.816} & 0.846 & \textbf{0.83}  & 0.355 & 57.368 \\ 
phi-3-mini      & 0.193 & 0.04  & \textbf{0.116} & \textbf{0.024} & 0.194 & 0.813 & 0.845 & 0.828 & \textbf{0.515} & 47.789 \\ 
qwen2.5-1.5b    & 0.179 & 0.029 & 0.097 & 0.017 & 0.182 & 0.802 & 0.847 & 0.823 & 0.427 & 46.0   \\ 
qwen2.5-3b      & 0.194 & 0.036 & 0.106 & 0.016 & 0.196 & 0.807 & 0.844 & 0.825 & 0.429 & 56.053 \\ 
smollm2-1.7b    & 0.185 & 0.037 & 0.11  & 0.022 & 0.197 & 0.81  & 0.846 & 0.827 & 0.437 & 50.842 \\ 
\bottomrule
\end{tabular}%
\caption{Fine-tune Result}
\label{tab:finetune_result}
\end{table*}

RQ3 examines the impact of Parameter-Efficient Fine-Tuning (LoRA) on model performance (Table~\ref{tab:finetune_result}). Fine-tuning most benefited ultra-lightweight models: Gemma-3-1b achieved its highest BERTScore F1 across all settings (+2.3~pp over zero-shot), confirming that the smallest models gain the most from domain-specific adaptation.

However, fine-tuning introduced significant regressions in models that performed well under other strategies. OLMo-2-1b exhibited the most notable decline, with its BERTScore F1 dropping by $-$4.4~pp and compliance falling by $-$13.9~pp relative to zero-shot. Phi-3-mini's compliance regressed by $-$24.3~pp from its few-shot peak, suggesting that fine-tuning on limited domain data can override the effective in-context calibration these models had already achieved.

\subsection{Results RQ4: To what extent do SLMs exhibit semantic diversity when proposing solutions to open-ended architectural requirements?}

\begin{figure*}
    \centering
    \includegraphics[width=0.5\linewidth]{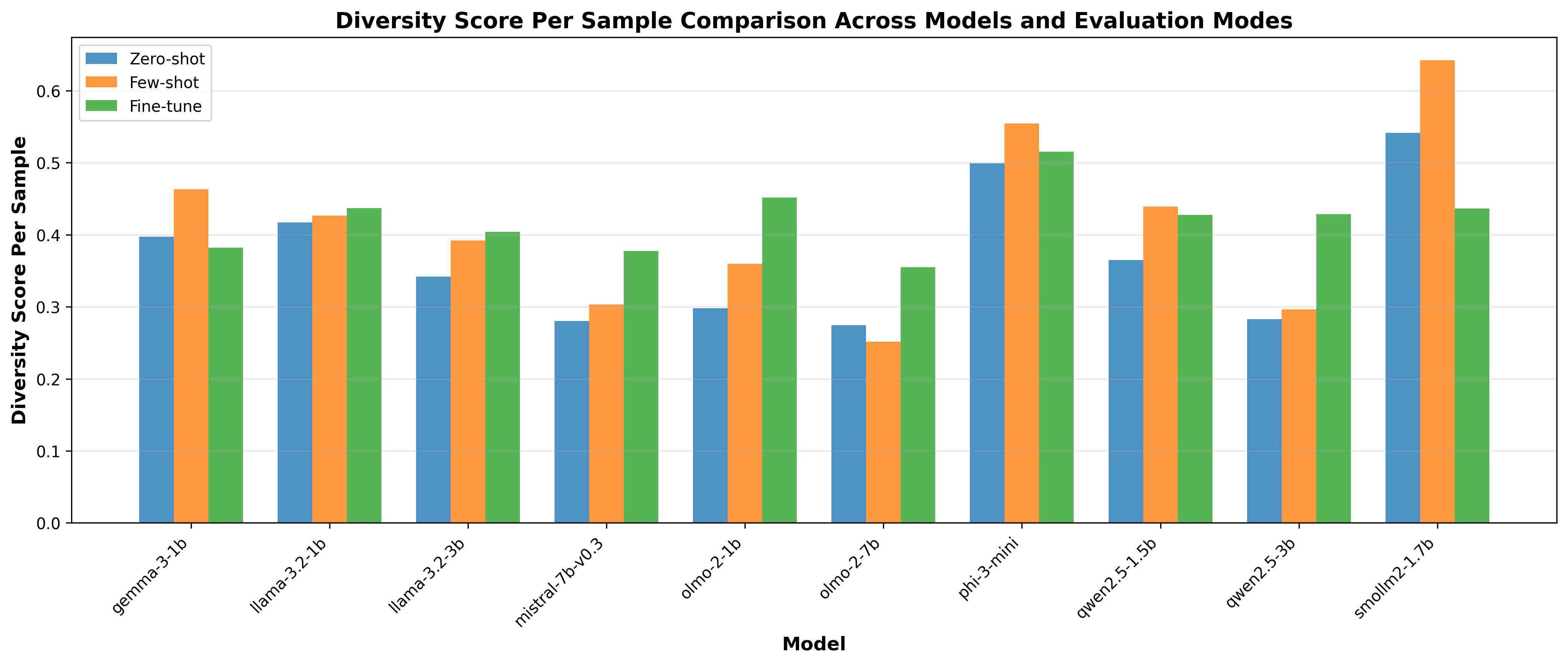}
    \caption{Diversity Score}
    \label{fig:diversity_score}
\end{figure*}

RQ4 reports the Semantic Diversity Score alongside Compliance Score across all three experimental settings (Tables~\ref{tab:zeroshot_result}--\ref{tab:finetune_result}, Figure~\ref{fig:diversity_score}).

In zero-shot, an inverse pattern emerges between diversity and compliance: the model with the highest diversity also exhibited the lowest compliance among sub-2B models, while the two least diverse models achieved the highest compliance scores in the cohort. This suggests that in small off-the-shelf models, high diversity often signals stochastic variance rather than productive exploration.

Under few-shot prompting, select models achieved the desirable outcome of simultaneous diversity and compliance gains (e.g., SmolLM2-1.7b: +14.6~pp compliance alongside +18.7\% relative diversity increase), while others saw divergent trends where diversity gains came at the expense of compliance.

After fine-tuning, diversity scores converged across most models into a narrow band (0.38--0.45), with Phi-3-mini and OLMo-2-7b as outliers above and below this cluster, respectively.

\begin{figure*}[t]
    \centering
    \begin{minipage}[b]{0.49\linewidth}
        \centering
        \includegraphics[width=\linewidth]{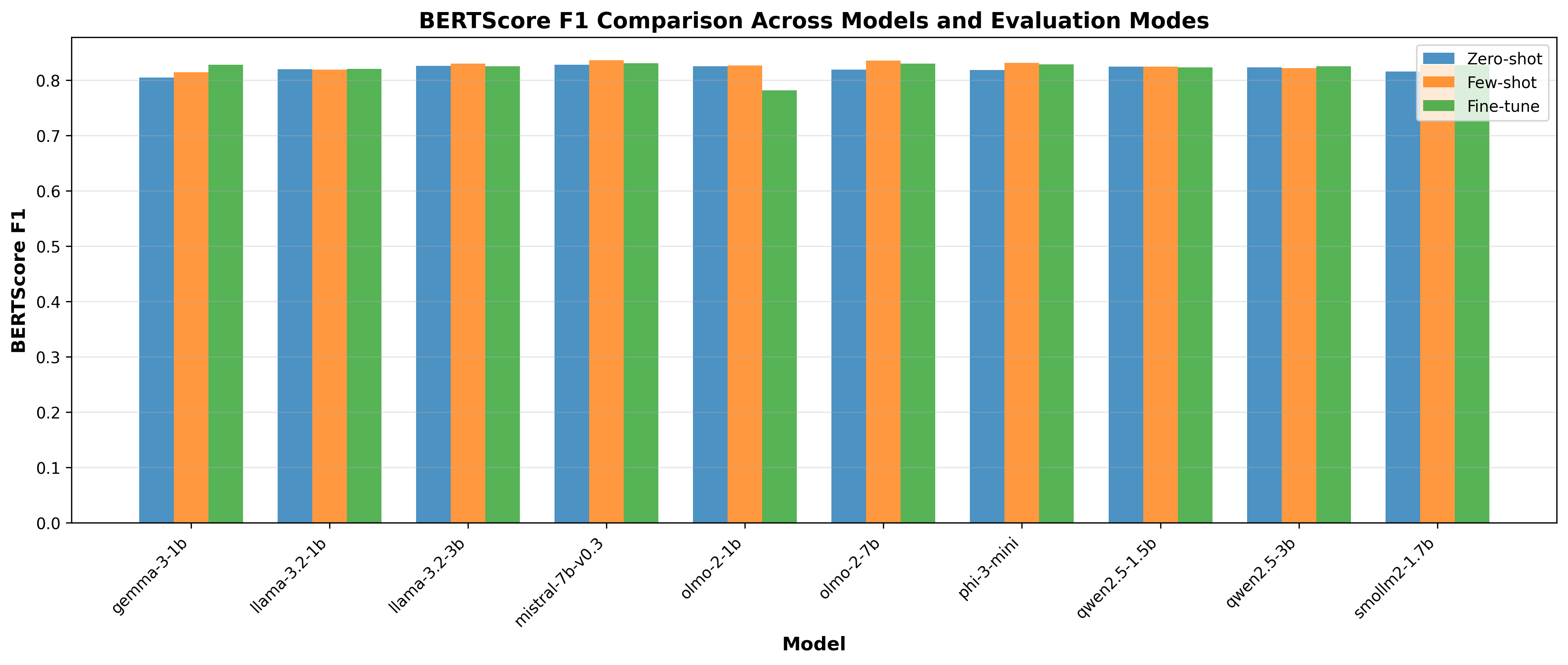}
        \caption{BERTScore f1}
        \label{fig:bertscore_f1}
    \end{minipage}
    \hfill
    \begin{minipage}[b]{0.49\linewidth}
        \centering
        \includegraphics[width=\linewidth]{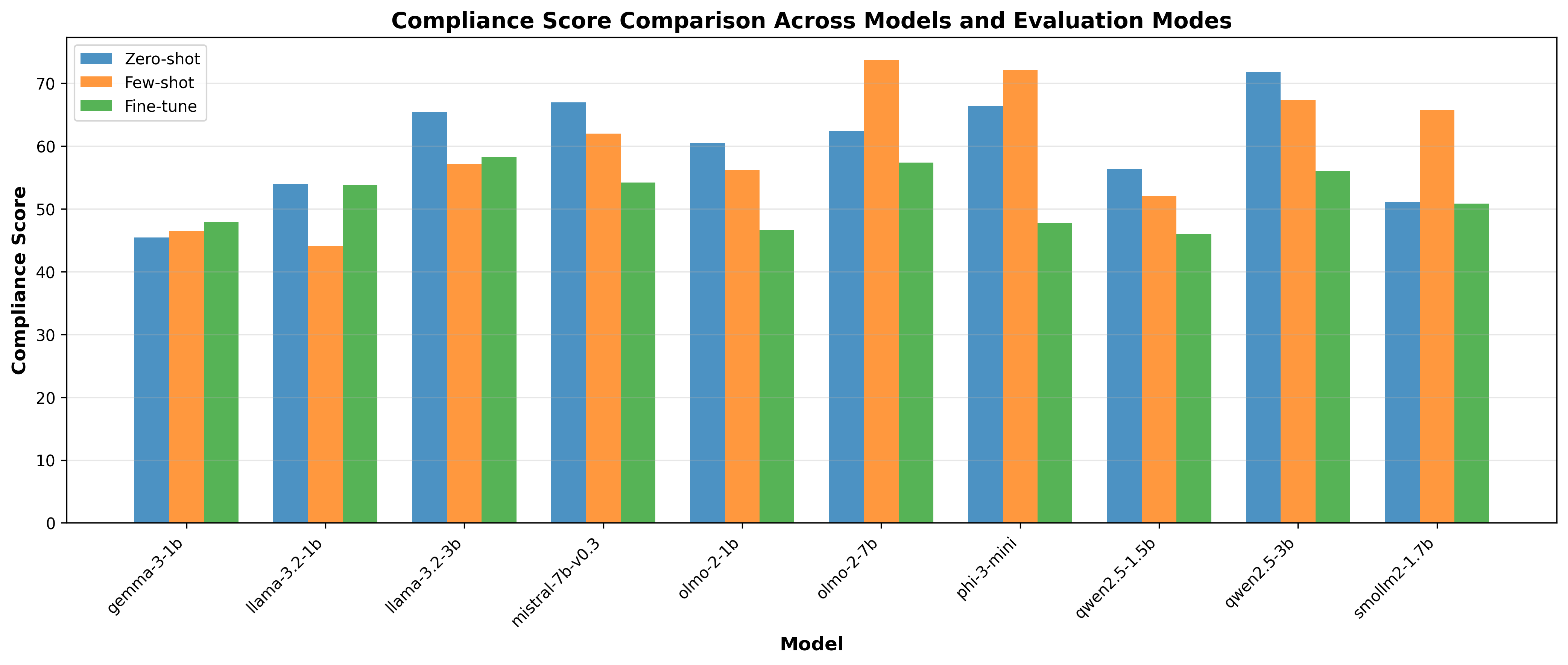}
        \caption{Compliance Score}
        \label{fig:compliance_score}
    \end{minipage}
\end{figure*}

\section{Discussion}
This study provides evidence that Small Language Models (SLMs) can be effectively leveraged to generate Architectural Design Decisions, challenging the assumption that massive cloud-based models are a prerequisite for this task. Below, we address each research question by examining the outcomes of our experiments and drawing conclusions regarding the efficacy of using SLMs for ADR generation. 

\subsection{How capable are off-the-shelf SLMs in generating semantically accurate and architecturally compliant design decisions? $(RQ_{1})$}

Yes, but capability is strictly tied to model scale.

Our zero-shot experiments reveal that models above the 3B-parameter threshold, such as Mistral-7b-v0.3 and Qwen2.5-3b, demonstrate a strong innate ability to propose valid architectural solutions, achieving high compliance scores without any task-specific training (Figure~\ref{fig:compliance_score}). This contradicts earlier assumptions that only massive proprietary Large Language Models can handle such reasoning tasks autonomously.

However, this capability is not universal. We identified a clear reasoning gap in the sub-2B parameter class (Figure~\ref{fig:compliance_score}). Models like Gemma-3-1b and Smollm2-1.7b, while capable of generating text with high lexical overlap, frequently failed to align with architectural best practices, evidenced by their low compliance score. Notably, this divergence between lexical similarity and architectural correctness is visible when contrasting Figures~\ref{fig:bertscore_f1} and~\ref{fig:compliance_score}: sub-2B models achieve competitive BERTScore F1 values (0.805--0.825) yet substantially lower compliance (45.4\%--56.3\%), confirming that textual fluency does not imply sound architectural reasoning. This suggests that for ultra-lightweight models, the pre-training data alone is insufficient to instill the complex decision logic required for SA.

\subsection{Does In-Context Learning (Few-Shot Prompting) improve the architectural compliance and semantic quality of SLMs? $(RQ_{2})$}

In-Context Learning is an effective calibration mechanism for select model classes, but its benefits are not universal. Providing just two examples enabled efficient models like Phi-3-mini to approach or match 7B-class baselines, boosting its compliance score from 66.4\% to 72.1\% - surpassing few-shot Mistral-7b (62.0\%).

This result indicates that the constrained context windows of these efficient models are not a bottleneck. Instead, the examples serve as a critical calibration signal, helping the model to adopt the correct tone and structural rigor of an ADR. For practitioners, this implies that few-shot prompting is a cost-effective deployment strategy for models with sufficient capacity and short context windows such as Phi-3-mini and OLMo-2-7b, as it yields state-of-the-art results without the computational overhead of fine-tuning.

However, ICL is not universally beneficial: Llama-3.2-1b and Mistral-7b-v0.3 both experienced compliance drops under few-shot prompting, suggesting that for models with already-strong zero-shot priors or those whose attention mechanism does not efficiently leverage in-context 
examples, the addition of examples introduces noise rather than a useful calibration signal.

\subsection{To what extent does domain-specific Fine-Tuning maximize architectural compliance in resource-constrained models? $(RQ_{3})$}

Fine-tuning is essential for the smallest models but optional for larger ones. 

Our results highlight a nuanced trade-off. For models lacking innate architectural intuition like Gemma-3-1b, fine-tuning yielded the strongest BERTScore improvement (0.805 to 
0.828), suggesting that weight updates help encode structural patterns even if compliance gains remain modest and do not generalize to all 1B-class models.

However, for capable architectures like Mistral-7b or Olmo-2-7b, fine-tuning yielded diminishing returns, producing results comparable to or slightly lower than the few-shot approach. Furthermore, the regression observed in OLMo-2-1b serves as a cautionary tale: over-optimizing small models on narrow datasets can degrade their general reasoning capabilities. Thus, fine-tuning should be reserved for ultra-low-resource scenarios where models under 3B parameters are required.

Similarly, Phi-3-mini's compliance regressed from 72.1\% under few-shot to 47.8\% after fine-tuning, reinforcing that models which already perform well via prompting are particularly susceptible to over-specialization on the narrow training set.

\subsection{To what extent do SLMs exhibit semantic diversity when proposing solutions to open-ended architectural requirements? $(RQ_{4})$}

While the tension between output diversity and generation quality is a known behavior across generative models generally, our results show this trade-off manifests with particular severity in SLMs: high diversity in small models more often signals hallucination than creative exploration, in contrast to larger models where diversity can coexist with high compliance. 

We observed a crucial distinction between "productive exploration" and "stochastic variance." In zero-shot settings, the smallest models exhibited the highest diversity scores but the lowest compliance. This suggests that their "diverse" outputs were largely incoherent or factually inconsistent.

In contrast, capable models like Mistral-7b exhibited lower diversity paired with higher compliance, indicating convergent reasoning - the model consistently identified the optimal architectural pattern across multiple runs. However, we found that In-Context Learning could successfully balance this trade-off. For select models such as Phi-3-mini, few-shot prompting 
successfully increased both diversity and compliance simultaneously (diversity: $0.499 \rightarrow 0.554$, compliance: $66.4\% \rightarrow 72.1\%$), while OLMo-2-7b achieved the largest compliance gain ($62.4\% \rightarrow 73.6\%$) with a modest reduction in diversity ($0.274 \rightarrow 0.252$), suggesting that examples can guide productive exploration rather than random generation in models with sufficient capacity. 

\section{Threats to validity}

In this section, we discuss the potential threats to the validity of our study and the employed mitigation strategies.

\textbf{Internal Validity:}
Prompt engineering bias may favor specific model families. We mitigated this through standardized, neutral prompt templates across all models, ensuring fair baseline comparison of instruction-following capabilities. Data leakage poses the risks that ADRs may have appeared in pre-training corpora; we rigorously separated training/testing splits to prevent immediate leakage. The use of 4-bit NF4 quantization, while aligned with resource-constrained deployment goals, may introduce precision loss affecting observed performance ceilings.

\textbf{External Validity:}
Our ADR dataset, while standard, represents a limited subset of SA domains and may not cover enterprise-scale distributed systems or legacy modernization scenarios. Model selection faces challenges from rapid AI development; we mitigated this by selecting 10 representative SLMs (1B-7B parameters) with diverse training methodologies to ensure broad coverage of the current SLM landscape.

\textbf{Construct Validity:}
The LLM-as-a-Judge mechanism for Architectural Compliance Score may harbor biases favoring verbose or stylistically similar answers. We implemented a strict scoring rubric focusing on technical correctness rather than style, with moderate, setting-dependent correlation between Compliance Score and BERTScore, confirming they capture complementary rather than redundant quality dimensions. High semantic diversity scores present interpretation challenges; we addressed this by analyzing diversity alongside compliance to distinguish productive exploration from incoherent hallucination. Additionally, as the judge model (Gemini-2.5-Flash) may have been pre-trained on data including the same open-source ADR repositories used in this study, a potential familiarity bias cannot be fully excluded, which may inflate compliance scores uniformly across all evaluated models. We also acknowledge the absence of a human-expert baseline: the Compliance Scores were not validated against independent grading by professional architects, which would provide stronger evidence of the judge's alignment with human expert judgment. Future work should incorporate such a human calibration study to further substantiate the reliability of the automated evaluation.

\textbf{Conclusion Validity:}
Generative model non-determinism may yield outlier results. We generated three distinct sequences per test input using Nucleus Sampling to ensure reliability and reduce random seed variation impact. For semantic metrics like BERTScore with compressed dynamic ranges, we interpreted values contextually based on recent literature confirming that small deltas correspond to significant qualitative performance shifts.

\section{Conclusion and Future Work}

\subsection{Conclusion}

This study presents SLM-ArchBench, a multi-dimensional evaluation framework that systematically probes the architectural reasoning capabilities of SLMs in generating ADRs. Through empirical investigation of 10 state-of-the-art SLMs, we establish that a clear capability boundary exists around the 3B-parameter threshold, with models like Mistral-7b-v0.3 achieving compliance scores exceeding 66\% in zero-shot settings while sub-2B models struggle to generate architecturally sound decisions autonomously. Few-Shot Learning is a highly effective calibration mechanism for select mid-sized models with short context windows, enabling models like Phi-3-mini and OLMo-2-7b to approach or exceed 7B-class baselines using merely two examples, as evidenced by Phi-3-mini's compliance improvement from 66.4\% to 72.1\%. Parameter-Efficient Fine-Tuning provides the strongest BERTScore improvements for ultra-lightweight models (1B) but does not reliably improve compliance and risks regression in certain architectures, while yielding diminishing returns or performance degradation in larger architectures. Our semantic diversity analysis reveals that high output variance in off-the-shelf small models often correlates with hallucination rather than productive exploration, while Few-Shot Learning successfully balances architectural correctness with valid solution space exploration. These findings establish actionable deployment guidelines:

\begin{itemize}
    \item \textbf{7B models} perform best under zero-shot or selective few-shot strategies and should avoid fine-tuning, which yields diminishing returns or risks performance degradation.
    \item \textbf{Select 3B--7B models} with short context windows (e.g., Phi-3-mini, OLMo-2-7b) benefit most from Few-Shot Learning, achieving state-of-the-art compliance without fine-tuning overhead.
    \item \textbf{1B models} may benefit from fine-tuning for semantic accuracy, though compliance gains are not guaranteed and architecture-specific regression is possible.
\end{itemize}

\noindent Together, these guidelines provide practitioners with evidence-based guidance for deploying sustainable, locally hosted architectural assistants in Software Engineering 2.0 environments.

\subsection{Future Work}

While this study establishes a rigorous baseline for evaluating SLMs in architectural reasoning, several directions warrant further investigation. Future work should extend the benchmark to encompass broader architectural tasks including system decomposition, quality attribute trade-off analysis, and architecture refactoring recommendations \cite{saip} to validate whether the identified parameter thresholds generalize beyond ADR generation. Incorporating multi-modal evaluation with visual architectural representations (UML, C4) and long-context scenarios would assess models' ability to bridge abstract design and concrete implementations \cite{palombo2026generate} \cite{chen2023autokg}. Investigating interactive refinement protocols where architects provide corrective feedback would inform the design of effective human-in-the-loop architectural assistants for SE 2.0 environments \cite{saip}. Finally, integrating comprehensive energy consumption metrics alongside reasoning performance \cite{h4} and evaluating model robustness under ambiguous requirements would provide critical insights for responsible production deployment of sustainable architectural assistants.

By pursuing these research directions, the software engineering community can advance toward truly effective AI-assisted architectural design-systems that augment human expertise rather than replace it, respect resource and privacy constraints, and maintain the rigorous reasoning standards required for sustainable software evolution in the SE 2.0 era.

\section{Code Availability}
The source code developed for this study and results can be found at \href{https://github.com/LeoTTTPhat/SLM-ArchBench}{https://github.com/LeoTTTPhat/SLM-ArchBench}

\section{Acknowledgment}
We acknowledge Ho Chi Minh City University of Technology (HCMUT), VNU-HCM for supporting this study.

\bibliographystyle{IEEEtran}
\bibliography{references}

\end{document}